# Detection of a Small Scale Cosmic Microwave Backround Anisotropy at 3.6 cm

E. A. Richards

University of Virginia and National Radio Astronomy Observatory[1]

E. B. Fomalont and K. I. Kellermann

National Radio Astronomy Observatory

R. B. Partridge

Haverford College

and

R. A. Windhorst

Arizona State University

---

## ABSTRACT

We report on the detection of a feature with negative radio flux in a sensitive, low resolution observation with the VLA. This morphologically peculiar feature is approximately $25'' \times 65''$ in size with a peak central amplitude of about −0.25 mK compared to the brightness temperature of the background sky at 3.6 cm. Within about $1'$ of this microwave decrement, we also found two radio quiet quasars, both at $z = 2.561$, with a projected physical separation of about 1 Mpc ($q_o = 0.5$, $h = 0.5$), suggestive of a possible galaxy cluster in the region. We are unable to account for this negative source by systematic phase fluctuations within our observations, sidelobe artifacts, or instrumental noise. We discuss possible physical origins of this microwave source, in particular the inverse-Compton scattering of the cosmic microwave background by hot gas in a distant cluster of galaxies.

*Subject headings:* (cosmology:) cosmic microwave background – cosmology: observations – galaxies: clusters: general – inter-galactic medium

## 1. Introduction

Clusters of galaxies are perhaps the largest gravitationally bound entities in the universe. As such, they are ideal laboratories for studying cosmological evolutionary scenarios in which structure formed through the action of gravity alone. The formation and subsequent evolution of galaxy clusters is highly sensitive to various cosmological parameters ($h_o, \Omega, \Lambda$) and the index, $n$, of the power spectrum of primordial mass perturbations, out of which all large-scale structure is believed to have formed. Thus, galaxy clusters provide potentially powerful probes of many aspects of modern cosmology (see Durret *et al.* 1994, for a review). Consequently much attention has been given to galaxy clusters in the past decade from both the observational and theoretical perspectives with the hope of using clusters to discriminate among the various cosmological models.

Galaxy clusters are known to possess deep gravitational potential wells filled with large amounts of ionized gas ($M_{gas} \sim 10^{13-14}$ $M_\odot$ or 0.1-0.01 $M_{cl}$). This intergalactic plasma gives rise to copious x-ray emission up to $L_x \sim 10^{44-45}$ erg/s in rich clusters. However, few high redshift ($z > 0.5$) clusters have been found in deep x-ray surveys. Several authors have interpreted this trend as a real effect due to cluster evolution, rather than an observational completeness problem (Edge *et al.* 1990; Henry *et al.* 1992). In essence, they argue that clusters of galaxies have undergone strong evolution with respect to redshift in the sense that there were fewer x-ray luminous clusters at earlier cosmological epochs, corresponding to $z \geq 0.5$. However, optical surveys out to $z \sim 1$ show no decline in the comoving number density of galaxy clusters (Dressler & Gunn 1990), implying that any apparent evolution may be confined to the x-ray emission of galaxy clusters. More recently, Luppino & Gioia (1995) have studied six rich clusters of galaxies with $z > 0.5$ and $L_x > 5 \times 10^{44}$ erg/s and have argued that there does appear to be an appreciable population of x-ray bright high redshift clusters. Thus the nature of cluster evolution remains unclear.

The principal obstacle in studying high redshift x-ray clusters of galaxies lies in the photon limited nature of x-ray astronomy. Although x-ray detectors have large fields of view and have high quantum efficiencies, the generally extended x-ray sources and the cosmological $(1 + z)^4$ surface brightness dimming (in addition to the possibly quite disadvantageous x-ray K-correction for steep spectrum galaxy clusters), dictates that most distant cluster observations are inherently photon starved. However, the same electrons which produce the thermal x-rays observed in clusters of galaxies also interact with the cosmic microwave background (CBR) through the inverse-Compton effect. This upscattering of CBR photons in frequency space depletes the CBR spectrum at frequencies shortward of 213 GHz, causing a microwave diminution towards the cluster (Sunyaev & Zel'dovich 1970). This CBR anisotropy, the Sunyaev-Zel'dovich (S-Z) effect, has now been observed in several rich clusters of galaxies at a range of redshift up to $z = 0.54$. (Birkinshaw & Hughes 1994; Herbig *et al.* 1995; Wilbanks *et al.* 1995; Carlstrom *et al.* 1996; Grainge *et al.* 1996 and refernces therein).

Unlike x-ray emission, the microwave surface brightness of a cluster as induced via the S-Z effect is virtually independent of distance and redshift (since the blackbody energy density, $u_{CBR} \propto (1 + z)^4$). Thus the S-Z signature of a cluster provides a powerful tool to search for distant clusters of galaxies, whose x-ray emission might otherwise be too faint to detect. Several authors have performed detailed analytical calculations and numerical simulations (Cole & Kaiser 1988, Schaefer & Silk 1988, Makino & Suto 1991; Bartlett & Silk 1993,

Markevitch *et al.* 1992, 1994) predicting both number counts of S-Z signals and the amplitude of small-scale fluctuations in the CBR due to a background of distant clusters for a variety of evolutionary scenarios and cosmolgical models. Generally, small scale S-Z fluctuations are expected in the amplitude range $\Delta T/T_r = 10^{-6} - 10^{-5}$, with $T_r$ the present temperature of the CBR, 2.726 K (Mather *et al.* 1994).

In the standard model for formation of galaxy clusters, large scale structure evolves from initial mass perturbations in the otherwise homogeneous primeval plasma. The cold dark matter scenario, in particular, assumes no preferred gravitational scale length and characterizes the evolution of cluster parameters by self-similar scaling with respect to cosmological redshift (see Kaiser 1986 for review). This simple theory predicts (for conventional values of $n$) that clusters were more numerous in the past for a given comoving volume element, that the cluster gas temperature was about the same as at present ($10^{7-8}$ K), and that clusters were less massive but physically smaller and hence denser in the past. As the S-Z effect depends only on the integral of the gas pressure along the line of sight, this theory predicts that the SZ signal of a given cluster could actually increase with redshift (Markevitch *et al.* 1994).

However, there are other tenable theories for the formation of large-scale structure. In the top-down or hot dark matter (HDM) cosmology, large scale structure forms first in the form of massive (M $\sim 10^{15}$ $M_\odot$ ) pancake-like structures which then fragment to form clusters and galaxies (Doroshkevich & Zel'dovich 1975). Bouchet *et al.* (1988) have also suggested that cosmic strings may play a crucial role in the formation of galaxy clusters and galaxies. The observation of clutser properties at large redshifts ($z > 1$) may help discriminate between these various scenarios.

## 2. The Radio Observations

We have used the VLA to search for CBR fluctuations on angular scales 6″- 80″ (Fomalont *et al.* 1993; Partridge *et al.* 1996). In the course of these observations, we detected an isolated, negative flux density feature in one field. Its properties suggest that it may be produced by the S-Z effect in a distant cluster of galaxies.

We used the VLA at 8.4 GHz from 1993 October through 1995 January for a total of 159 hours to observe a region of the sky free of bright sources, and centered at (J2000) $\alpha = 13^h12^m17.4^s$ and $\delta = +42° 38' 05''$. This field was chosen because it did not contain any radio sources with flux density, $S_{8.4GHz} \geq 0.5$ mJy and had been previously observed by the Hubble Space Telescope as part of the Medium Deep Survey (*e.g.*,Windhorst *et al.* 1995). Observations in both the C and D configurations of the VLA were made and combined. The shortest and longest antenna baselines were 0.035 km and 3.4 km respectively.

The synthesized beam of the combined data was about 6″ full width at half maximum (FWHM) and the field of view was 312″ to the primary beam half power points. Our observing scheme was to alternate between phase calibrator 1244+408 and our primary field with 3 minute and 27 minute integrations, respectively. Our flux density and polarization calibration (we observe both right and left circular polarizations simultaneously) was provided by observations of 3C286. The data was generally of high quality with residual phase fluctuations always less than 5° and correlator gain variations less than 3%.

After proper editing, weighting, and cleaning (see Partridge *et al.* 1996 for details), our full resolution image had a *rms* sensitivity of 1.5 $\mu$Jy, making it the most sensitive radio image obtained to date at any frequency or resolution. We also made lower resolution images by applying Gaussian tapers to the (u,v) data in order to downweight the longer baseline visibilities, effectively broadening the synthesized beam. Although these tapered images have a higher point to point *rms* noise, the microwave brightness temperature sensitivity is increased because of the increase in size of the resolution element ($T_{rms} \propto \theta^{-2}$).

For the purposes of searching for CBR inhomogeneities, we located all the positive radio sources greater than 7 $\mu$Jy (4.7 $\sigma$) in our full resolution map and subtracted their components from the (u,v) data, as described by Partridge *et al.* (1996) and Fomalont *et al.* (1993). All subsequent analysis was conducted on these residual images made with various tapers. These maps are free from sidelobe contamination from strong sources but were not CLEANed further.

## 3. Detection of a Negative Radio Feature

In the combined C + D array data with 6″ resolution, we discovered an extended source, *negative* in flux density, located approximately 25″ from the image phase center. At the –5.5 $\sigma$ level and with only 680 independent beams in our field of view, the feature has a greater than 99% chance of being a real source in the sky (assuming the image noise has Gaussian properties). This feature was also found in the independent C-array and D-array images alone, further proof to its reality. This microwave feature is also prominent at lower resolutions [1] where at 30″ resolution it has a peak amplitude of –13.9 ± 3.3 $\mu$Jy or –250 ± 60 $\mu$K. Although the shape of the source changes somewhat at lower resolutions as it begins to blend with positive residual emission in the map (i.e. real sources present in the image but below our 7 $\mu$Jy completeness limit and hence not subtracted from the data - see below for discussion), it is clearly extended and at 30″ resolution is 25″× 65″, oriented roughly in the N-S direction. Images of the negative microwave feature at 18″ and 30″ resolution are shown in figures 1 and 2, respectively.

---

[1] Although at 30″ resolution the decrement is only at the –4$\sigma$ level as compared to the –5.5$\sigma$ level in the full resolution map, the decreased number of independent beams (27 as compared to 680) in the lower resolution image gives the detection in this image virtually the same statistical significance.

### 3.1. Evidence that the Microwave Diminution is not an Instrumental Artifact

We performed a variety of tests to determine whether a feature with these properties could be produced by instrumental effects, such as the overlap of negative sidelobes of the discrete foreground sources in the field. We have checked this possibility by comparing the extent and amplitude of the feature in a deeply cleaned map (in which source sidelobes are much reduced) with the same parameters in our essentially uncleaned map. The angular position and extent were essentially unchanged, and the peak amplitudes were –7.9 $\mu$Jy and –8.2 $\mu$Jy for the cleaned and uncleaned maps, respectively (both at 6″ resolution). Corresponding figures for the lower (30″) resolution maps were –13.3 $\mu$Jy and –13.9 $\mu$Jy.

While bright (S ≥ 7 $\mu$Jy) sources and their sidelobes have been subtracted from our map, a multitude of faint sources remain as evidenced by the steep integral source count slope of $\mu$Jy radio sources of $\alpha \sim -1.3$ (Richards *et al.* 1996) in our field. Could the combined sidelobes of these weak sources produce a negative artifact of the amplitude that we observed? We tested this hypothesis by performing a Monte Carlo analysis of our observations. These simulations had exactly the same (u,v) coverage as our observations, as well as the same spatial distributions for all sources stronger than about 6 $\mu$Jy. Thus, we took all of the brighter radio sources found in our full resolution image at the 4 $\sigma$ level and above, and placed them on top of a faint population of weaker sources generated according to an empirically determined source count (Richards *et al.* 1996, Kellermann *et al.* 1996, Windhorst *et al.* 1993). After mapping these simulated data in 20 independent trials at 6″ resolution and removing the bright sources as we did with our actual data, no residual features $\leq -4.5\ \sigma$ were found, consistent with well behaved thermal noise centered on zero map flux. Because the sidelobe contribution to the residual signal in our maps is dominated by the 'brighter' sources stronger than 4 $\sigma$, only a few Monte Carlo trials were needed to demonstrate that sidelobe artifacts cannot account for the observed decrement (i.e., the large number of weak sources in the image whose spatial distribution we simulated as random cannot alone account for the negative flux feature).

We also investigated the possible effects of systematic phase errors on our images. Using Monte Carlo simulations like those described above, we introduced phase offsets of up to 15° into our raw visibilities. We searched these simulated images for negative features of similar magnitude to that seen in our obersved image. None were found $\leq -4.5\sigma$, where our signal is $-5.5\sigma$. This analysis indicates that our observed negative feature is a true decrement in the CBR.

## 4. Supporting Observations

HST observations of this radio field as part of the Medium Deep Survey (Windhorst *et al.* 1994) were made to limiting $3\sigma$ point source sensitivities of $m_v = 26.4$ and $m_I = 24.8$. The 21 radio/optical identifications are published elsewhere (Windhorst *et al.* 1995). Optical spectra obtained at the Multiple Mirror telescope (MMT) showed that two of the microJansky radio sources are quasars at identical redshifts of 2.561. Examination of the individual spectra of the two quasars argues against the possibility that they are gravitationally lensed as both the Ly$\alpha$ emission and the continua have significantly different shapes. However, because of the large spatial separation ($\theta \sim 100''$), the differential time delay along the two optical paths is expected to be quite large, on the order of 500 years (Turner *et al.* 1986). Thus it is conceivable that this hypothetically lensed quasar might have evolved over this period of time. The implied lensing mass in this scenario is about $10^{15}$ $M_\odot$ .

The likelihood of finding two quasars separated by less than 1 Mpc ($h_o = 0.5$, $q_o = 0.5$), with $m_v = 18.4$ and $m_v = 21.0$ at this redshift in a Poissonian distribution (using the quasar luminosity function of Kron *et al.* 1991) is much less than 1%. Clustering of quasars on scales of less than 10 Mpc has been known for at least a decade (Shank *et al.* 1984). The spatial distribution of quasars is believed to trace the underlying dark matter potential field of the universe (Fisher *et al.* (1990) and thus the presence of the quasar pair suggests the possible existance of a $z = 2.56$ cluster of galaxies which could produce our negative microwave feature. We note that the field of view of the HST WFPC1 is only about 1 Mpc at $z = 2.56$, somewhat comparable to the the Abell radius for a rich cluster. Although there is a group of 10-20 galaxies near the microwave decrement (see Windhorst *et al.* 1995), it is uncertain if they are physically associated with the $z = 2.561$ quasars or are merely foreground objects. The HST optical point source detection limits at this redshift are fairly modest, M $\sim$ -21 at rest-frame 230 nm and M $\sim$ -20 at rest-frame 170 nm and thus are not optimal for detecting high redshift galaxy cluster members.

## 5. Possible Physical Explanations

There are several mechanisms capable of producing fluctuations in the microwave sky at the arcminute scale (e.g., Partridge 1995 for review). However, only anisotropies arising from cosmic strings and the S-Z effect can produce isolated *negative* fluctuations. Cosmic string perturbations (Bouchet *et al.* 1989) are believed to arise from the relativistic propagation of spatial defects. Numerical simulations show that these fluctuations should appear as step-like discontinuities in microwave images, producing a net deficit of CBR photons in the wake of the cosmic string and a symmetric increase in temperature in the opposite direction. Our feature obviously does not fit this geometrical criterion.

Primary CBR fluctuations produced at $z \sim 1000$ by both the Sachs-Wolfe and the Doppler effect are expected to have a Gaussian distribution, with negative and positive fluctuations about $T_r$ equally present. However, observational limits on the amplitude of such primary fluctuations on comparable and somewhat larger angular scales (Readhead *et al.* 1989; Fomalont *et al.* 1993; Subrahmanyan *et al.* 1993 and Partridge *et al.* 1996) make it very unlikely that our $\Delta T/T_r \sim 10^{-4}$ negative feature is a primary fluctuation.

## 6. The Physical Parameters of this Cluster

If the the negative flux feature we observe is due to the S-Z effect in a galaxy cluster in our radio field, what constraints can other observations, particularly in x-rays, set on its properties? We assume that the cluster is at the same redshift as that of the $z = 2.561$ quasars to calculate its physical parameters. Sarazin (1986) shows that the S-Z decrement in a cluster of galaxies can be expressed as:

$$\Delta T = 0.085 \left(\frac{n_o}{10^{-3} cm^{-3}}\right) \left(\frac{T_g}{10^8 K}\right) \left(\frac{r_c}{0.25 Mpc}\right) \left(\frac{T_r}{2.7 K}\right) \times$$
$$\left(\frac{\Gamma(3\beta/2 - 1/2)}{\Gamma(3\beta/2)}\right) (1 + x^2)^{-3\beta/2 + 1/2} \ mK \qquad (1)$$

Here $r_c$ is the cluster core radius, $T_r$ is the present CBR temperature, $T_g$ is the gas temperature of the inter-cluster plasma, $n_o$ is the central electron density of the cluster medium, $\beta$ is the isothermal gas profile index, $\Gamma$ is the gamma function, and $x$ is defined as $r/r_c$. For an isothermal gas profile we can write the inter-cluster gas distribution as:

$$n_e = n_o \left[1 + \left(\frac{r}{r_c}\right)^2\right]^{-3\beta/2} \qquad (2)$$

For simplicity, we will ignore the elongation of the observed decrement, and treat it as arising from a spherical distribution producing a FWHM given by the average of the two axes. For rich clusters, a typical value of $\beta = 2/3$. Then, the FWHM of the microwave decrement is $a_l = 2\sqrt{3} \ r_c$. We can now express the central intracluster (ICM) pressure as :

$$n_o T_g = \left(\frac{2\sqrt{3}}{\sqrt{\pi}}\right) \left(\frac{\Delta T}{0.085 mK}\right) \left(\frac{0.25 Mpc}{a_l}\right) \left(\frac{2.7 K}{T_r}\right) \times 10^1 \ K cm^{-3} \qquad (3)$$

In the above equation, $\Delta T$ corresponds to the central value of the cluster decrement. For a given cosmology and redshift, the scale length, $a_l$, of the cluster can be derived from the angular size of the observed decrement and and the angular diameter-redshift relation. Thus, for a given cosmological model, the parameters of the feature we observe provide a measurement of the pressure integral through the galaxy cluster.

It is useful to calculate the range of expected x-ray fluxes from a galaxy cluster capable of producing the microwave diminution that we observe. We will then compare these expected x-ray fluxes with the sensitive *ROSAT HRI* image of the region provided by Hu and Cowie (1996). Using a Raymond-Smith plasma emission model which includes both x-ray line emission and free-free radiation with 30% solar heavy element abundance, we have calculated the x-ray luminosity (in the observed 0.1 - 2.4 keV band) for a range of cluster ICM temperatures, $T_c$, cosmologies (different values of $h$, $q_o$, and $\Lambda$), and $\beta$ parameters. We use the emission integral characterized as:

$$EI = 6.74 \times 10^{65} \left(\frac{n_o}{10^{-3}cm^{-3}}\right)^2 \left(\frac{a_l}{0.25 Mpc}\right)^3 \left(\frac{\Gamma(3\beta - 3/2)}{\Gamma(3\beta)}\right) \ cm^{-3} \qquad (4)$$

As a specific example, figure 3 gives the expected x-ray flux for for $q_o = 0.5, \Lambda = 0, h_o = 0.5$ and $\beta = 2/3$ as a function of the cluster temperature, $T_g$. Representative values derived from the observed $\Delta$T measurement for the cluster parameters in the above cosmology are $L_x \sim 2 \times 10^{44}$ ergs/s, $M_{gas} \sim 10^{13}$ $M_\odot$ , and $r_c = 0.1$ Mpc (for an assumed $T_g \sim 5$ keV).

Although no significant x-ray flux was found in the *ROSAT HRI* image within the expected core radius of the cluster ($r_c = 15''$ for $\beta = 2/3$) as measured from the microwave map, we derived a $3\sigma$ detection limit of $f_x \leq 2 \times 10^{-14}$ $ergs \ cm^{-2} \ s^{-1}$ . After convolution with the redshifted *ROSAT HRI* passband and after applying the K correction assuming the Raymond-Smith spectral model, it appears that a cluster of the x-ray luminosity we have calculated would have been just at the detection limit of the Hu and Cowie *HRI* observation for any cosmology or intracluster gas model considered here. Thus, the *HRI* image does not rule out a cluster at $z = 2.56$, but does eliminate the possibility that our negative feature is produced by a background cluster with $z$ less than 1.5.

## 7. Confrontation with Models

The promise of using CBR observations to detect high redshift galaxy clusters through their S-Z signature has been discussed by a number of authors (e.g., Markevitch *et al.* 1994 and references therein). The detection of substantial numbers of clusters at redshifts greater than 0.5 is essential to accurately determine the nature of cluster evolution which will itself place stringent constraints on various models for galaxy and cluster formation and evolution. We turn now to more specific, model-dependent predictions.

Ceballos & Barcons (1994) have recently investigated the expected contribution of the S-Z effect to anisotropies in the microwave sky at both centimeter and millimeter wavelengths. The results of Ceballos & Barcons rely heavily on the work of Edge *et al.* (1990) who analyzed an all sky x-ray survey by the *Einstein* and *EXOSAT* observatories which they claim to be complete to a flux level of $1.7 \times 10^{-11}$ $ergs$ $cm^{-2}$ $s^{-1}$ and to a depth of $z = 0.1$. Using the strong negative evolution function found by Edge *et al.* and a simple parameterization of the x-ray gas profile within individual clusters, Ceballos & Barcons extrapolated this x-ray cluster luminosity function to redshifts up to $z \sim 3$. With this formalism, they conclude that there should be no fluctuations at 3.6 cm in the microwave sky due to the S-Z effect above $\Delta T/T = 10^{-8}$ except those in known x-ray bright galaxy clusters. In short they have calculated the expected S-Z signal under the assumption of strong negative cluster evolution. Our observed signal of $\Delta T/T = 10^{-4}$ is a full four orders of magnitude greater than the result predicted by Ceballos & Barcons. Our observations would seem to contradict the strong negative evolution in the cluster population they assumed.

Another recent treatment of the S-Z problem with respect to the anisotropy expected from a background of high redshift clusters is given by Bartlett & Silk (1994). Using the Press-Schecter formalism to construct the mass function for clusters and using simple self-similar scaling relations to model the evolution of the ICM, Bartlett & Silk predict *rms* CBR fluctuations on arcminute scales for a flat universe (i.e., $\Omega = 1$) and for $n = $ -1 and -2. Because the authors predict *rms* amplitudes of CBR fluctuations rather than S-Z number counts and amplitudes, we compare the model to CBR measurements at $1'$ resolution as given by Partridge *et al.* 1996 ($\Delta T/T = 1.4 \pm 1.2 \times 10^{-5}$). Our observations rule out all scenarios with a bias parameter, b greater than 1 and $n$ less than -1. Our results are consistent with the Bartlett & Silk prediction of $\Delta T/T = 8 \times 10^{-6}$ in an $\Omega = 1$, $n = $ -1 universe.

Finally, we compare our results to the predictions of Markevitch *et al.* (1994). They calculate the number counts of S-Z sources as a function of $\Omega$, the redshift of cluster formation, $z_{max}$ and the index of the power spectrum of initial density perturbations, $n$. They normalize their number counts by using the observed number of local x-ray clusters (Vikhlinin *et al.* 1994), and evolving those back in time in a self-similar fashion for various choices of cosmological parameters. Thus they predict specific number counts of S-Z sources in the sky for each unique combination of $\Omega$, $z_{max}$, and $n$. Their most optimistic models ($\Omega$ $= 0.3$, $z_{max} = 5$, and $n = 1$, predict at most one S-Z source of amplitude $\Delta T/T = 10^{-4}$ per degree of sky. The VLA CBR experiments at 3.6 cm cover a total area of roughly 0.017 deg$^2$ in two fields (another deep VLA field has recently been imaged at 3.7 cm and *no* negative sources are detected at less than –9 $\mu$Jy, Fomalont *et al.* 1996) and thus imply a sky density of S-Z sources of $\sim 60 \pm 40$ per deg$^2$ at the level of $\Delta T/T = 10^{-4}$. If our observed microwave decrement is due to the S-Z effect in a distant cluster of galaxies then it is marginally consistent with the $\Omega = 0.3$, $z_{max} = 5$, and $n = 1$ model of Markevitch *et al.* but manifestly contradicts all other scenarios considered by those authors. We can also place an upper limit on the number of S-Z sources in the sky from our observations. At the 90% confidence level, there are fewer than 230 S-Z sources per $deg^2$ with S less than –10 $\mu$Jy. Not suprisingly this does not place an interesting limit on any of the models considered by Markevitch *et al.*

However, drawing conclusions from small number statistics such as ours can be misleading. One must also be careful in using Poisonian statistics in describing the non-Gaussian spatial distribution of clusters although in this respect we are helped somewhat by the fact that the galaxy cluster clustering length is of order 10 Mpc (Bahcall & West 1992) whereas the field of view of our radio observations at $z = 1$ is about 3 Mpc. We also make note that this is not the first CBR anistropy found at arcminute scales. Meyers *et al.* (1993) discovered a signal in their RING experiment with an amplitude of $+466 \pm 90$ $\mu$K (probably a foreground source). Also a group at the Ryle telescope (Jones *et al.*) have found a significant negative CBR fluctuation in a field with two known high redshift quasars.

## 8. Conclusions

We have shown that our observations of a $\Delta T = -0.25$ mK microwave 'dip' at 8.4 GHz are consistent with a Sunyaev-Zel'dovich signal in a cluster of galaxies at $z > 1.5$. The observation of two quasars with identical redshifts in the field suggest that this cluster candidate is possibly at $z = 2.561$. The existence of a rich cluster of galaxies at such a large look-back time is problematic for hierarchical CDM theory where large scale structure formed late in the evolution of the universe. However, rather than make generalizations about structure formation from one observation, we instead emphasize the power of S-Z observations of high redshift clusters to discriminate among the various cosmological models. For instance, if our observed microwave feature is due to a rich cluster of galaxies at high redshift, then cosmological models with n < 1 are ruled out (Markevitch *et al.* 1994) as the strong negative evolution of clusters in these models makes the detection of even one rich cluster at such large look-back times extremely unlikely.

At 2 mm this particular source should be appreciably brighter (S ∼ –2 mJy) if it is due to the to the S-Z effect. At 0.8 mm where the positive S-Z signal is at a maximum, the expected flux density is about 1.6 mJy. This should be readily observable in a reasonable exposure at several of the new mm and sub-mm observatories (e.g., SCUBA). X-ray observations with the AXAF observatory would be useful to confirm this possible detection of a galaxy cluster. Near-infrared observations in this region would be useful to search for optical members of this high redshift cluster candidate as well.

## 8.1. Acknowledgments

We are indebted to J. Uson for helpful comments and criticisms concerning the possible observational errors in this project. In addition, we are grateful to C. Sarazin and M. Markevitch who provided both stimulating conversations and helpful suggestions in the preparation of this work. We also thank E. Hu and L. Cowie for allowing us access to their x-ray data prior to its publication. This work is in part based on observations with the NASA/ESA Hubble Space Telecope, obtained at the Space Telescope Science Institue, and was supported by NASA grant GO-2684.03-87A from STScI, which is operated by the Association of Universities for Research in Astronomy, Inc., under NASA contract NAS5-26555. Optical observations were obtained in part at the Multiple Mirror Telescope Observarory, a joint facility of the University of Arizona and the Smithsonian Institution. RBP acknowledges support from NSF grant AST-9320049 to Haverford College.

## 9. References

Anantharamaiah, K. R., Despande, A. A., Radhakrishnan, V., Ekers, R. D., & Goss, W. M. 1991, IAU Symposium 131, Radio Interferometry: Theory, Techniques and Applications

Bartlett, J. G. & Silk, J. 1994, ApJ , 423, 12

Bennett, C. L., Banday, A. J., Gorski, K. M., Hinshaw, G., Jackson, P., Keegstra, P., Kogut, A., Smoot, G. F., Wilkinson, D. T., Wright, E. L. 1996, ApJ, 464L

Birkinshaw, M., Hughes, J. P. 1994, ApJ, 420, 1

Bouchet, F. R., Bennett, D. P., & Stebbins, A. 1988, Nature, 335, 410

Carlstrom, J. E., Joy, M., Grego, L. 1996, ApJ, 456L, 75L

Ceballos, M. T., Barcons, X. 1994, ASTRO-PH9405052 (http://xxx.lanl.gov/archive/astro-ph)

Cole, S. & Kaiser, N. 1988, MNRAS, 233, 637

Doroshkevich, A. G. & Zeldovich, Ia. B. 1975, Astrophysics & Space Science, 35, 55

Durret, F., Mazure, A., Tran Thanh Van, J. 1994, Proceedings of the 29th Recontre De Moriond: Clusters of Galaxies

Edge, A. C., Stewart G. C., Fabian, A. C. & Arnauld, K. A. 1990, MNRAS, 245, 559

Fomalont, E. B., Partridge, R. B., Lowenthal, J. D., Windhorst, R. A. 1993 ApJ, 404, 8

Grainge, K, Jones, M, Pooley, G., Saunders, R., Baker, J., Haynes, T. & Edge, A. 1996, MNRAS, 278L, 17

Gunn, J. E. 1990, in Space Telescope Science Institue Symposium Series, "Clusters of Galaxies"

Henry, J. P., Gioia, I. M., Maccaro, T., Morris, S. L., Stocke, J. T. & Wolter, A. 1992, ApJ, 386, 408

Herbig, T., Lawrence, C. R., Readhead, A. C. S. & Gulkis, S. 1995, ApJ, 449L, 5L

Vikhlinin, A., Forman W., Jones, C., & Murray, S. 1995, ApJ, 451, 564

Kaiser, N. 1986, MNRAS, 222, 323

Kellermann, K. I., Fomalont, E. B., Richards, E. A, Partridge, R. B., Windhorst, R. A. 1996, in preparation

Kron, R., Bershady, M. A., Munn, J. A., Smetanka, J. J., Majewski, S., Koo, D. C. 1991, in Astronomical Society of the Pacific Conference Series: The Space Distribution of Quasars

Luppino, G. A, & Gioia, I. M. 1995, ApJ, 445L, 77L

Makino, N. & Suto, Y. 1993, ApJ, 405, 1

Markevitch, M., Blumenthal, G. R., Forman, W., Jones, C. & Sunyaev, R. A. 1992, ApJ, 395, 326

Markevitch, M, Blumenthal, G. R., Forman, W., Jones, C., Sunyaev, R. A. 1994, ApJ, 426, 1

Mather, J. C. *et al.* 1994, ApJ, 420, 439

Ostriker & Cowie 1981, ApJ, 243L, 127L

Partridge, R. B 1995, 3K, Cambridge University Press

Partridge, R. B., Richards, E. A., Fomalont, E. B., Kellermann, K. I., & Windhorst, R. A. 1996, ApJ, submitted

Readhead, A. C. S., Lawrence, C. R., Myers, S. T., Sargent, W. L. W., Hardebeck, H. E., & Moffet, A. T. 1989, ApJ, 346, 566

Richards, E. A., Fomalont, E. B., Kellermann K. I., Partridge, R. B., Windhorst, R. A 1995 in IAU 179: Extragalctic Radio Sources

Sarazin C. L. 1988, X-ray Emission from clusters of Galaxies, Cambridge University Press

Schaeffer & Silk 1988, ApJ, 333, 509

Subrayamanyan, R., Ekers, R. D., Sinclair, M., & Silk, J. 1993, MNRAS, 263, 416

Sunyaev, R. A., & Zel'dovich, Y. B. 1970, Astrophys. Space Sci., 7, 3

Turner, E. L., Gunn, J. E., Schneider, D. P., Burke, B. F., Hewitt, J. N. 1986, Nature, ,

321, 142

Wilbanks, T. M., Ade, P. A. R., Fischer, M. L., Holzapfel, W. L. & Lange, A. E. 1994, ApJ, 427L, 75L

Windhorst, R. A, Fomalont, E. B., Kellermann, K. I., Partridge, R. B., Richards, E. A., Franklin, B. E., Pascarelle, S. M., & Griffiths, R. E. 1995 Nature, 375, 471

Windhorst, R. A. *et al.* 1994, AJ, 107, 3

– 22 –## 10. Figure Captions

1. The extended negative feature found in our tapered $18''$ resolution map is shown. The peak emission is in the north lobe of the structure and is about -10 $\mu$Jy. The double-lobed nature of the source us unusual and may suggest that it is a dynamically unrelaxed or merging cluster.

2. The microwave decrement shown at $30''$ resolution. The peak emission here is about -14 $\mu$Jy. The low level positive radio emission is a blend of real sources, below our completeness level and hence not removed and noise.

3. The shaded area represents the flux levels that the *ROSAT HRI* image 3 $\sigma$ flux limit. Also plotted is the expected flux from a galaxy cluster with the gas pressure that we measure from our microwave image. We plot the expected x-ray flux as a function of the assumed gas temperature.

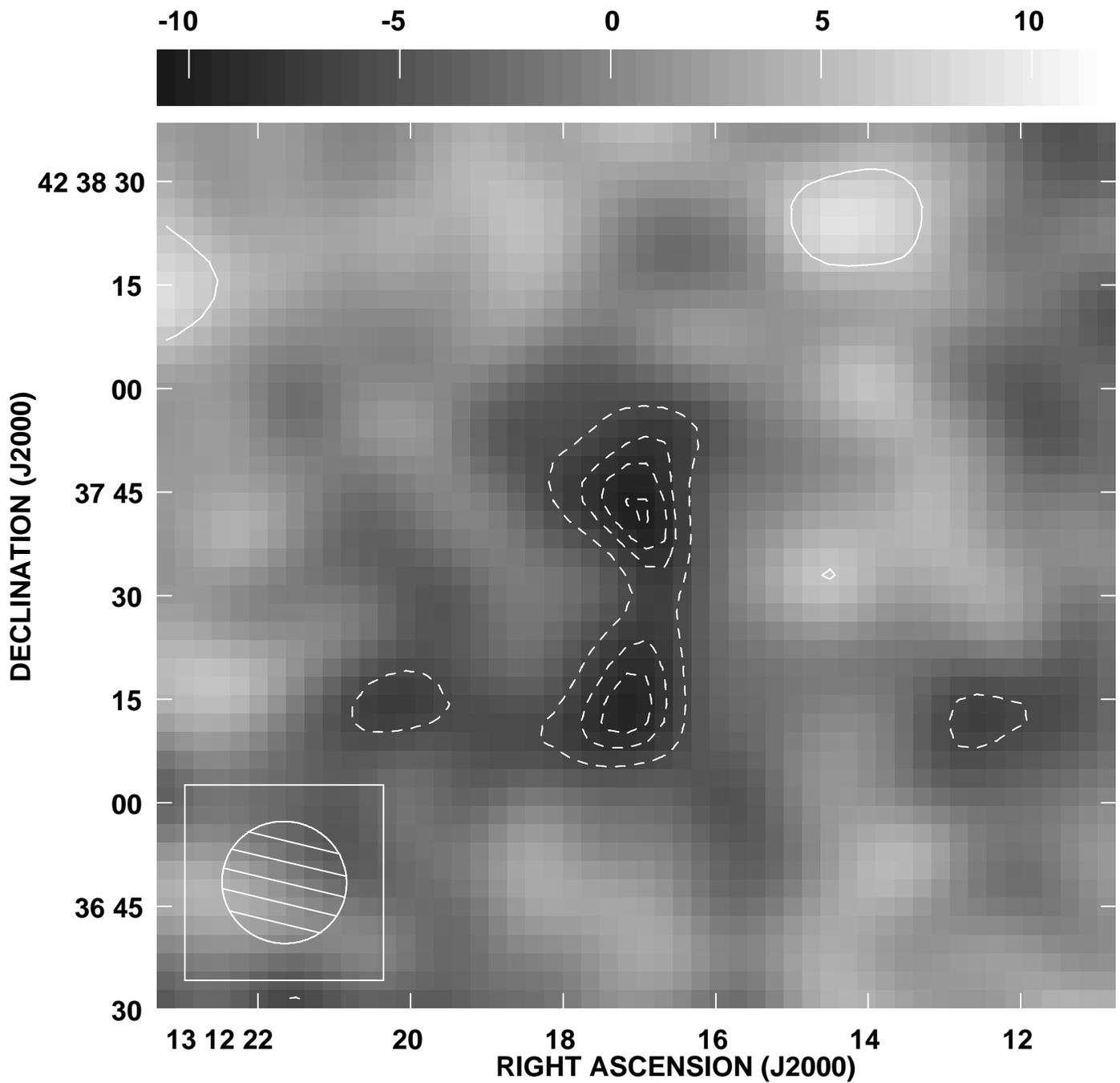

Cont peak flux =  1.1769E-05 JY/BEAM
Levs =  2.0000E-06 * (  -4.90, -4.40, -3.90,
 -3.00, 3.000)

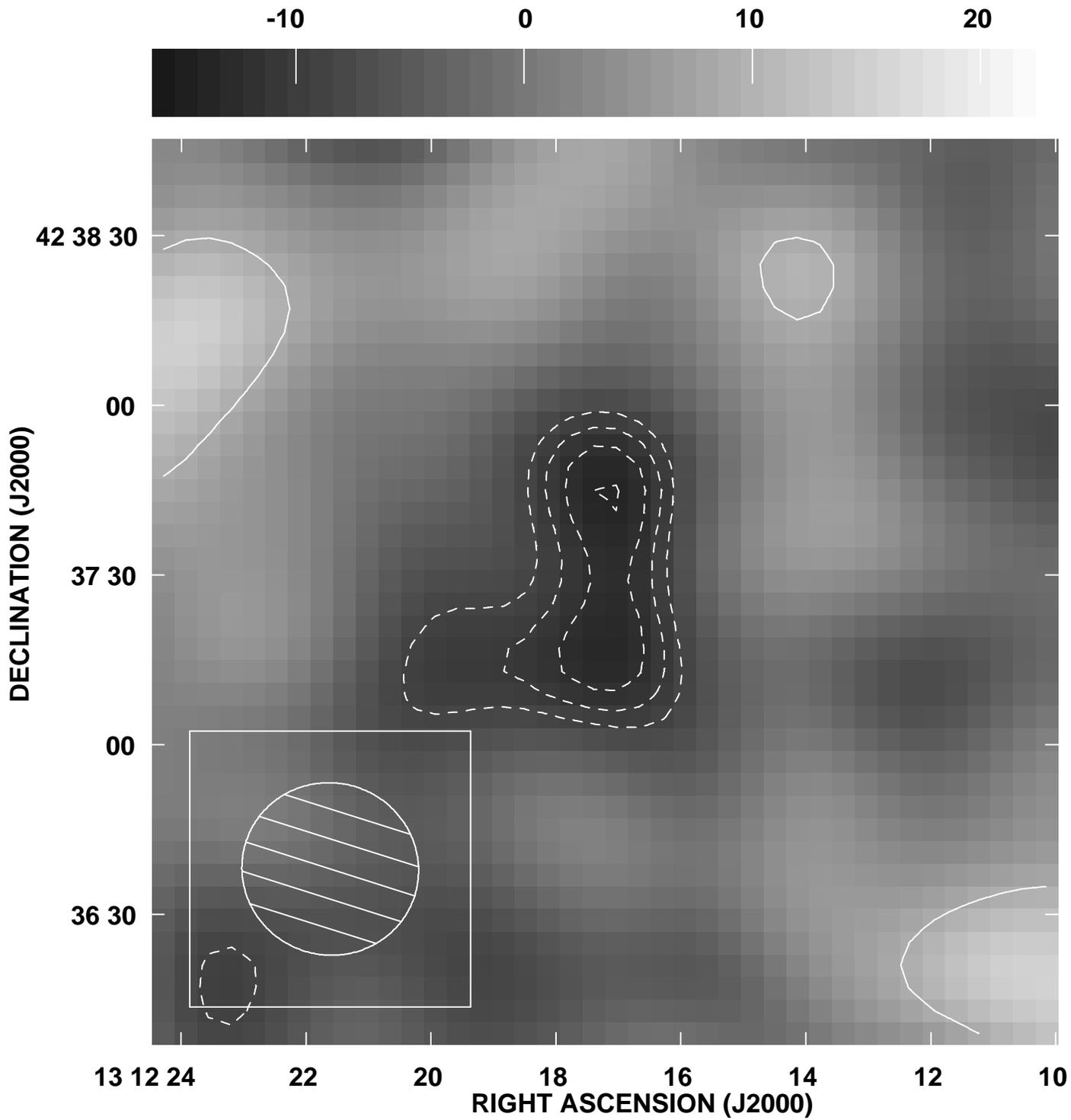

Cont peak flux = 2.2878E-05 JY/BEAM
Levs = 3.3000E-06 * ( -4.15, -3.65, -3.15, -2.65, 2.650)

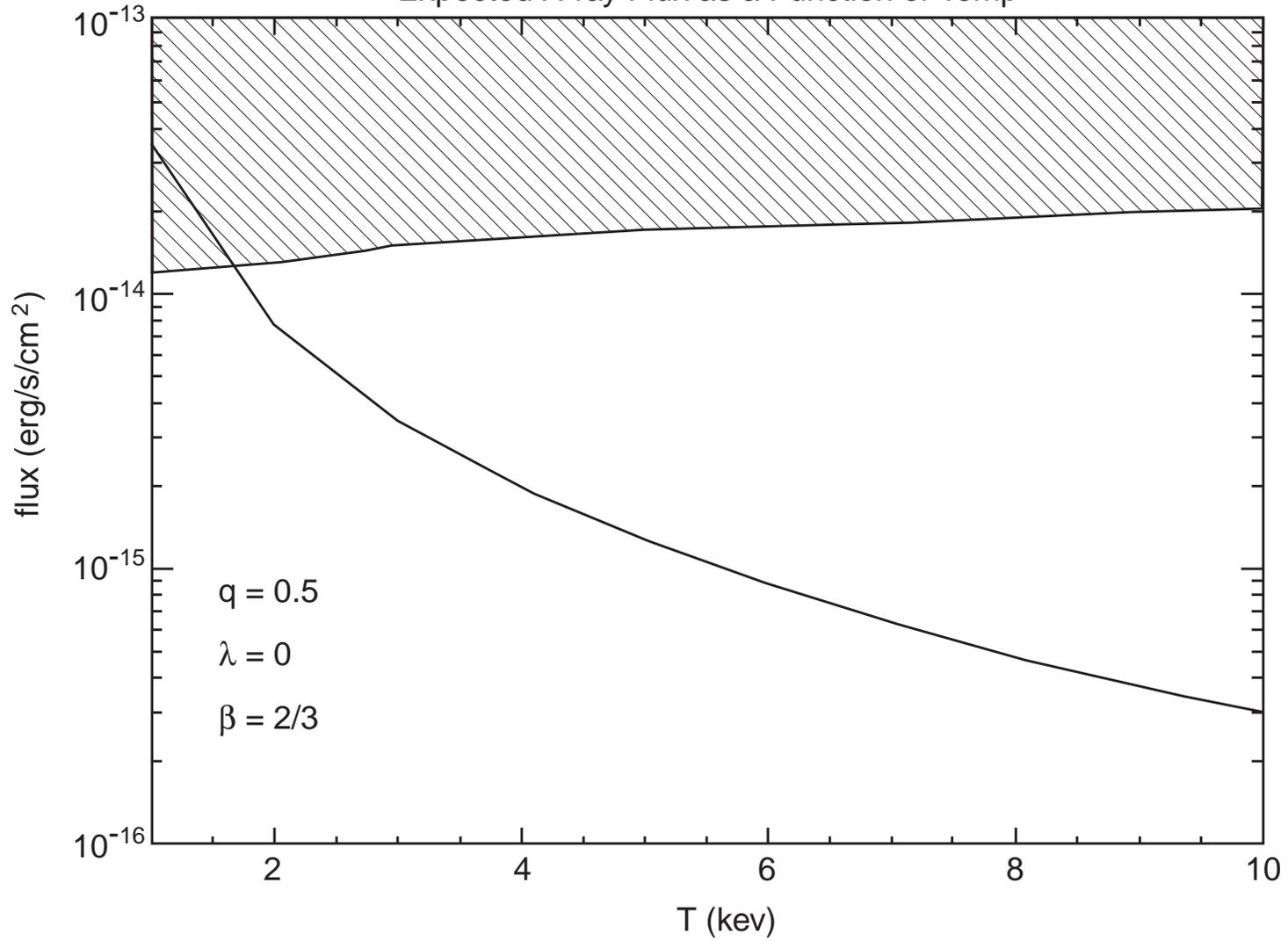